\documentclass[
superscriptaddress,
nofootinbib,
twocolumn,
 amsmath,amssymb,
 aps,
pra,
notitlepage,
longbibliography
]{revtex4-1}

\usepackage{dcolumn}
\usepackage{bm}
\usepackage{epsfig}
\usepackage{graphicx}
\usepackage{latexsym}
\usepackage{amsfonts}
\usepackage{setspace}
\usepackage{graphicx}
\usepackage{amsmath}
\usepackage{verbatim}
\usepackage{color}
\usepackage{SIunits}
\usepackage{hyperref}
\usepackage{physics}

\newcommand{\be}{\begin{equation}}
\newcommand{\ee}{\end{equation}}
\newcommand{\ba}{\begin{array}}
\newcommand{\ea}{\end{array}}
\newcommand{\bqa}{\begin{eqnarray}}
\newcommand{\eqa}{\end{eqnarray}}

\begin{document}

\title{InGaP quantum nanophotonic integrated circuits with $1.5\%$ nonlinearity-to-loss ratio}

\author{Mengdi Zhao} 
\affiliation{Holonyak Micro and Nanotechnology Laboratory, University of Illinois at Urbana-Champaign, Urbana, IL 61801 USA}
\affiliation{Department of Physics, University of Illinois at Urbana-Champaign, Urbana, IL 61801 USA}
\affiliation{Illinois Quantum Information Science and Technology Center, University of Illinois at Urbana-Champaign, Urbana, IL 61801 USA}
\author{Kejie Fang} 
\email{kfang3@illinois.edu}
\affiliation{Holonyak Micro and Nanotechnology Laboratory, University of Illinois at Urbana-Champaign, Urbana, IL 61801 USA}
\affiliation{Illinois Quantum Information Science and Technology Center, University of Illinois at Urbana-Champaign, Urbana, IL 61801 USA}
\affiliation{Department of Electrical and Computer Engineering, University of Illinois at Urbana-Champaign, Urbana, IL 61801 USA}

\begin{abstract} 

Optical nonlinearity plays a pivotal role in quantum information processing using photons, from heralded single-photon sources, coherent wavelength conversion to long-sought quantum repeaters. Despite the availability of strong dipole coupling to quantum emitters, achieving strong bulk optical nonlinearity is highly desirable. Here, we realize quantum nanophotonic integrated circuits in thin-film InGaP with a record-high ratio of $1.5\%$ between the single-photon nonlinear coupling rate ($g/2\pi=11.2$ MHz) and cavity-photon loss rate . We demonstrate second-harmonic generation with an efficiency of $71200\pm10300\%$/W in the InGaP photonic circuit and photon-pair generation via degenerate spontaneous parametric down-conversion with an ultrahigh rate exceeding 27.5 MHz/$\mu$W---an order of magnitude improvement of the state-of-the-art---and a large coincidence-to-accidental ratio up to $1.4\times 10^4$. Our work shows InGaP as a potentially transcending platform for quantum nonlinear optics and quantum information applications. 

\end{abstract}

\maketitle

\section{Introduction}

Optical nonlinearity is indispensable for a number of quantum information protocols using photons. A figure-of-merit characterizing nonlinear quantum systems is the ratio of single-photon nonlinear coupling rate ($g$) and photon loss rate ($\kappa$), which roughly measures how fast quantum information can be manipulated before it is lost. Deterministic quantum logic gates can be realized in the strong coupling regime (i.e., $g/\kappa>1$), for example, via cavity-quantum electrodynamics systems. Surprisingly, weak optical nonlinearity marked by $g/\kappa<1$, which is typical for systems with bulk nonlinear susceptibilities, might enable alternative protocols for quantum communication and computation, besides its wide use for (heralded) single-photon generation. For example, when accompanied with ancillary coherent states, weak optical nonlinearity might enable quantum non-demolition measurement of flying photons, which is useful for relaying quantum information \cite{nemoto2004nearly,munro2005weak,xia2016cavity}. Moreover, continuous-variable cluster states synthesized from the squeezed vacuum have been proposed to implement fault-tolerant quantum computing when the squeezing exceeds certain thresholds \cite{menicucci2014fault,fukui2018high,baragiola2019all}. These protocols, while avoiding the strong coupling regime, nevertheless might still need a nonlinearity-to-loss ratio $g/\kappa$ beyond those typically achievable with bulk nonlinear optical structures.

One method to enhance photon-photon interaction via bulk nonlinear susceptibilities is to make optical micro-cavities and circuits such that the photons confined in wavelength-scale structures interact accumulatively throughout their lifetime. The key to this approach is a thin-film material platform that yields, simultaneously, a large nonlinear mode coupling and low optical losses, thus large $g/\kappa$.  In particular, photonic integrated circuits have been developed in a growing number of thin-film materials with substantial second-order nonlinear susceptibility ($\chi^{(2)}$), such as gallium arsenide (GaAs) \cite{kuo2014second,chang2019strong} and aluminum gallium arsenide  \cite{mariani2014second,may2019second}. However, these III-V materials are associated with intrinsic optical losses due to, for example, two-photon or sub-bandgap absorption at near-infrared and telecom wavelengths \cite{michael2007wavelength}. Recently, wide bandgap materials, including aluminum nitride \cite{guo2017parametric,bruch201817} and lithium niobate \cite{luo2018highly,zhang2019broadband,lu2020toward,ma2020ultrabright}, are emerging among the leading nonlinear photonic platforms. With microcavities of ultrahigh quality factors realized, these material systems have yielded second-harmonic generation with an unprecedented efficiency \cite{lu2020toward} and ultra-bright heralded single-photon sources \cite{ma2020ultrabright}. 

Here, we explore another III-V material, indium gallium phosphide (InGaP), and demonstrate its unique strength for quantum nonlinear optics and quantum information applications. InGaP has a large second-order nonlinear susceptibility  ($\chi^{(2)}\approx 220$ pm/V \cite{ueno1997second}) comparable to GaAs and a sizable bandgap of about 1.9 eV which helps suppress two-photon absorption at telecom wavelengths. Previous studies of InGaP nonlinear photonics focuses on using its Kerr nonlinearity in waveguides and photonic crystal cavities for frequency combs and optical parametric oscillators \cite{dave2015dispersive,marty2021photonic}. We explore its substantial second-order nonlinearity and develop an insulator-on-top fabrication process to realize high-$Q$ quasi-phase-matched microring resonators with strongly coupled 1550 nm and 775 nm wavelength modes in thin-film InGaP. We observed a nonlinear coupling rate of $g/2\pi=11.2$ MHz and a nonlinearity-to-loss ratio $g/\kappa=1.5\%$ in InGaP microring resonators, both of which are the highest among all demonstrated nonlinear photonic platforms \cite{lu2020toward,ma2020ultrabright,kuo2014second,logan2018400,bruch201817}. Further, we show photon-pair generation via degenerate spontaneous parametric down-conversion with an ultrahigh rate exceeding 27.5 MHz/$\mu$W, which is an order of magnitude higher than previously reported record in periodically-poled lithium niobate microrings \cite{ma2020ultrabright}, and a large coincidence-to-accidental ratio up to $1.4\times 10^4$.

\section{Device design and simulation}
The devices deployed in this study are waveguide-coupled doubly-resonant microring resonators. The microring resonator supports fundamental 1550 nm and second-harmonic 775 nm resonances coupled via the second-order optical nonlinearity. The interaction between the two resonances can be described by the Hamiltonian, $\hat{H}=\hbar (g\hat{a}^{\dag 2}\hat{b}+g^*\hat{a}^2\hat{b}^{\dag})$, where $\hat{a}(\hat{a}^\dag)$ and $\hat{b}(\hat{b}^\dag)$ are the annihilation(creation) operators for the fundamental and second-harmonic resonance, respectively, and $g$ is the single-photon mode coupling coefficient. For disordered InGaP used in this study with a zinc blende crystal structure, its second-order nonlinear optical susceptibility only has the component $\chi^{(2)}_{xyz}$, leading to a mode coupling coefficient given by \cite{luo2019optical}
\begin{equation}
\begin{aligned}
	g & = \sqrt{\frac{\hbar\omega_a^2\omega_b}{8\epsilon_0}}\frac{\int d\mathbf{r} \chi^{(2)}_{xyz} \sum\limits_{i\neq j\neq k}E_{ai}^*E_{aj}^*E_{bk}}{\int d\mathbf{r} \epsilon_{\textrm{r}a}|\mathbf{E}_a|^2\sqrt{\int d\mathbf{r} \epsilon_{\textrm{r}b}|\mathbf{E}_b|^2}},
\end{aligned}
\label{eq:coupling g}
\end{equation}
where $\omega_{a(b)}$ is the angular frequency of resonance $a$($b$), $\mathbf{E}_{a(b)}$ is the modal electric field, $\epsilon_0$ is the vacuum permittivity, and $\epsilon_{\textrm{r}a(b)}$ is the relative permittivity at 1550 nm (775 nm). The $z$-direction is defined to be perpendicular to the device plane.

\begin{figure}[!htb]
\begin{center}
\includegraphics[width=1\columnwidth]{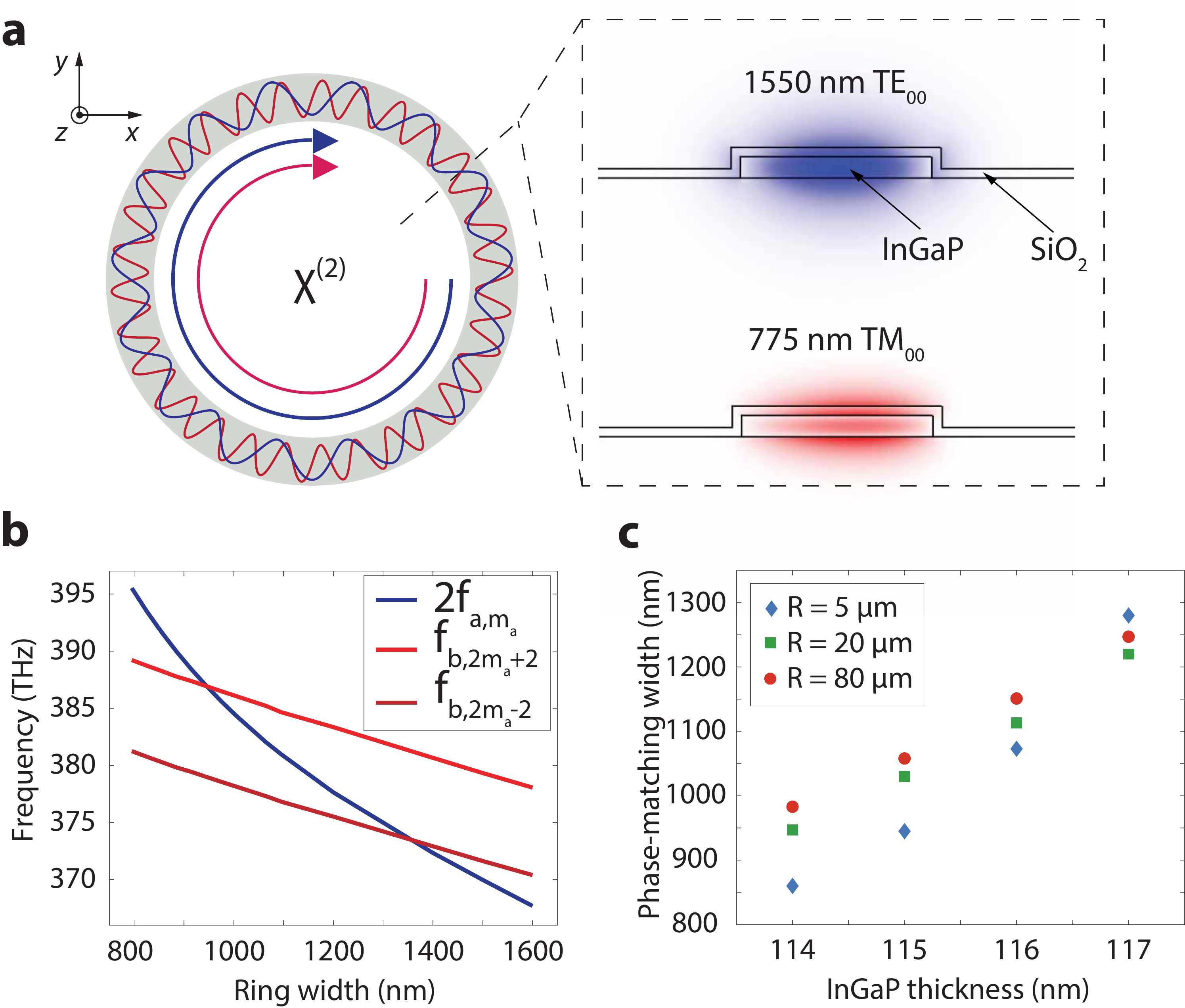}
\caption{\textbf{a}. Schematic plot of a $\chi^{(2)}$ microring resonator. Inset shows the electric field of the 1550 nm $\mathrm{TE_{00}}$ resonance ($E_r$ component) and 775 nm $\mathrm{TM_{00}}$ resonance ($E_z$ component) at the cross section of the microring.  \textbf{b}. Dispersion curve of the fundamental mode with azimuthal number $m_a=36$ and second-harmonic modes with azimuthal number $m_b=2m_a\pm 2$. The microring has a thickness of 115 nm and $R=5~\mu$m. \textbf{c}. The width of the microring satisfying the phase-matching condition $m_b=2m_a+2$ between the 1550 nm and 775 nm resonances.}
\label{fig:field profile&PM}
\end{center}
\end{figure}

To optimize $g$, and thus the device-level nonlinearity, modes $a$ and $b$ need to be phase-matched and have a substantial field overlap defined by the inter-modal integral in Eq. \ref{eq:coupling g}. We choose the fundamental transverse-electric mode (TE$_{00}$) and fundamental transverse-magnetic mode (TM$_{00}$) for the 1550 nm and 775 nm resonances (Fig. \ref{fig:field profile&PM}a), respectively, taking advantage of their dominant in-plane (for TE$_{00}$) and out-of-plane (for TM$_{00}$) electric field components. The $\chi^{(2)}_{xyz}$ nonlinear susceptibility imposes the energy conservation and quasi-phase-matching condition \cite{yang2007enhanced} between the two traveling-wave resonances for maximal coupling (Appendix \ref{App:gDerivation}), i.e.,
\begin{equation}
    2\omega_a=\omega_b, \quad |2m_a-m_b|=2,
\label{eq:PM}
\end{equation}
where $m_{a(b)}$ is the azimuthal number of resonance $a(b)$.

Because of the strong dispersion of InGaP with a refractive index of $n=3.12$ at 1550 nm and $n=3.41$ at 775 nm \cite{tanaka1986refractive}, the quasi-phase-matching condition can only be satisfied in thin InGaP microrings with a high-aspect-ratio cross section. Fig. \ref{fig:field profile&PM}b shows the dispersion of the TE$_{00}$ mode with $m_a=36$ and the TM$_{00}$ modes with $m_b=2m_a\pm2$ for the 115 nm thick microring with a center radius of $R=5~\mu$m. The crossing points of the dispersion curves thus yield the ring width, around 945 nm and 1360 nm, respectively, satisfying Eq. \ref{eq:PM}. A thin layer (50 nm) of silicon dioxide on top of the suspended InGaP microring is included in the simulated structure, in accordance with the fabricated device (see Section \ref{Sec:fab}). 
Using $\chi^{(2)}_{xyz}=220$ pm/V \cite{ueno1997second}, we numerically find a coupling coefficient $g/2\pi=10.90 (5.96)$ MHz between the 1550 nm TE$_{00}$ and 775 nm TM$_{00}$ resonances of the $5~\mu$m microring satisfying the quasi-phase-matching condition $m_b=2m_a+2 (m_b=2m_a-2)$. It is worth noting that $g$ scales approximately as $R^{-0.5}$ and thus smaller microring resonators will yield a larger nonlinear coupling, though photon lifetimes might be sacrificed due to excess scattering and bending losses.
Fig. \ref{fig:field profile&PM}c reveals that the quasi-phase-matching condition is sensitive to the thickness of InGaP films, requiring uniform InGaP films for optimized devices.

\vspace{-1cm}

\section{Device fabrication}\label{Sec:fab}

\begin{figure*}[!htb]
\begin{center}
\includegraphics[width=1.5\columnwidth]{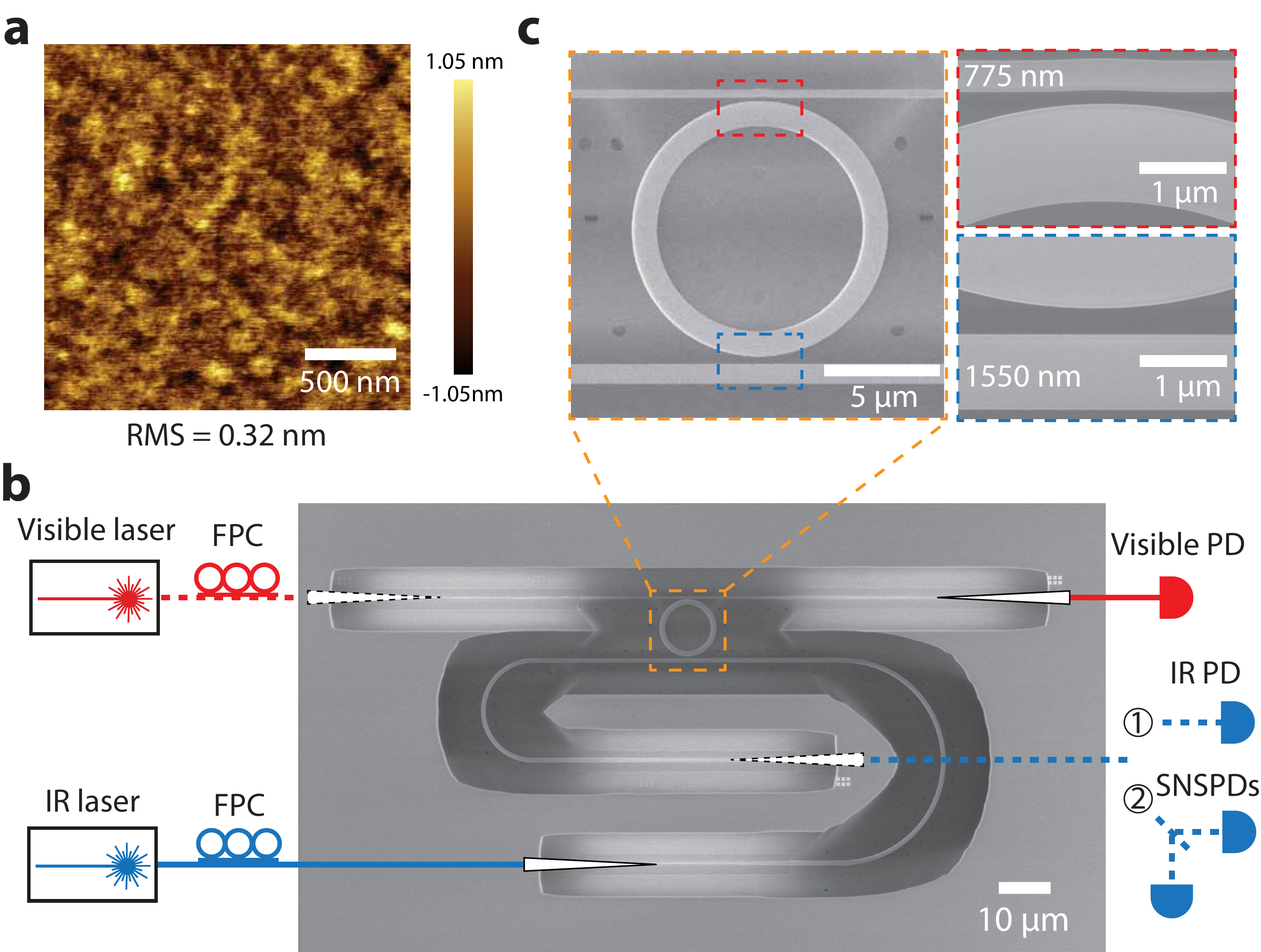}
\caption{\textbf{a}. Surface morphology of the 115 nm thick InGaP-on-GaAs film in a $2~\mu$m by $2~\mu$m region measured by AFM. The root mean square (RMS) of the surface roughness is 0.32 nm. \textbf{b} and \textbf{c}. Scanning electron microscopy images of the fabricated device and schematic of the measurement setup. Second harmonic generation and photon pair generation are measured via the solid and dashed path, respectively.  IR: infrared, FPC: fiber polarization controller, PD: photodetector, SNSPD: superconducting nanowire single-photon detector.}
\label{fig:setup}
\end{center}
\end{figure*}

The devices are fabricated from 115 nm thick (measured by scanning electron microscopy) disordered InGaP films grown on GaAs substrate (2 degree off-cut toward the [110]) by metal-organic chemical vapor deposition (T = 545 C, V/III = 48, precursors: trimethylindium, trimethylgallium and PH$_3$). The photoluminescence and X-ray measurements reveal a composition of disordered In$_{0.48}$Ga$_{0.52}$P with a 1.92 eV bandgap. The root-mean-square surface roughness of the InGaP thin film is about 0.32 nm measured by atomic-force microscopy (AFM) (Fig. \ref{fig:setup}a). The device pattern is defined using 150 keV electron beam lithography and 150 nm negative tone resist hydrogensilsesquioxane (HSQ). A 20 nm thick layer of silicon dioxide is deposited on InGaP via plasma-enhanced chemical vapor deposition (PECVD) to enhance the adhesion of HSQ. The pattern is transferred to InGaP via inductively coupled plasma reactive-ion etch (ICP-RIE) using Cl$_2$/CH$_4$/Ar gas mixture with a selectivity of InGaP: HSQ: PECVD SiO$_2$ = 240: 90: 80. After a short buffered oxide etch to remove the residual oxide (both HSQ and PECVD oxide), a layer of 50 nm thick silicon dioxide is deposited on the chip via atomic layer deposition. A second electron beam lithography and subsequent ICP-RIE using CHF$_3$ gas are applied to pattern etch-through holes in the silicon dioxide layer for undercut of the InGaP device. Finally, the InGaP device is released from the GaAs substrate using citric acid-based selective etching \cite{uchiyama2006fabrication}. The suspended InGaP device is mechanically anchored to the silicon dioxide membrane (see Appendix \ref{App:fabflow} for additional information and a schematic of the fabrication process).

Fig. \ref{fig:setup}b and c show scanning electron microscopy images of the fabricated device. The microring resonator is coupled to two bus waveguides for transmitting the 1550 nm and 775 nm wavelength light, respectively. The 1550 nm wavelength straight waveguide is 800 nm wide and separated from the ring by 400 nm. It decouples from the 775 nm TM$_{00}$ microring resonance because of the tight field confinement of the latter and the large ring-waveguide gap. On the other hand, the 775 nm wavelength pulley waveguide is 280 nm wide with a wrap angle of 6 degrees and a ring-waveguide gap of 250 nm. It decouples from the 1550 nm TE$_{00}$ microring resonance because of the significantly different mode index of the 1550 nm TE$_{00}$ mode in the narrow waveguide and the wide microring. In this way, we are able to separately transmit 1550 nm and 775 nm wavelength light in the two waveguides without crosstalk and independently control the external quality factors of the 1550 nm and 775 nm microring resonances. The bus waveguides are connected to adiabatically tapered couplers for transmitting TE- or TM-polarized light between the photonic integrated circuits and single-mode optical fibers.

\begin{figure*}[!htb]
\begin{center}
\includegraphics[width=2\columnwidth]{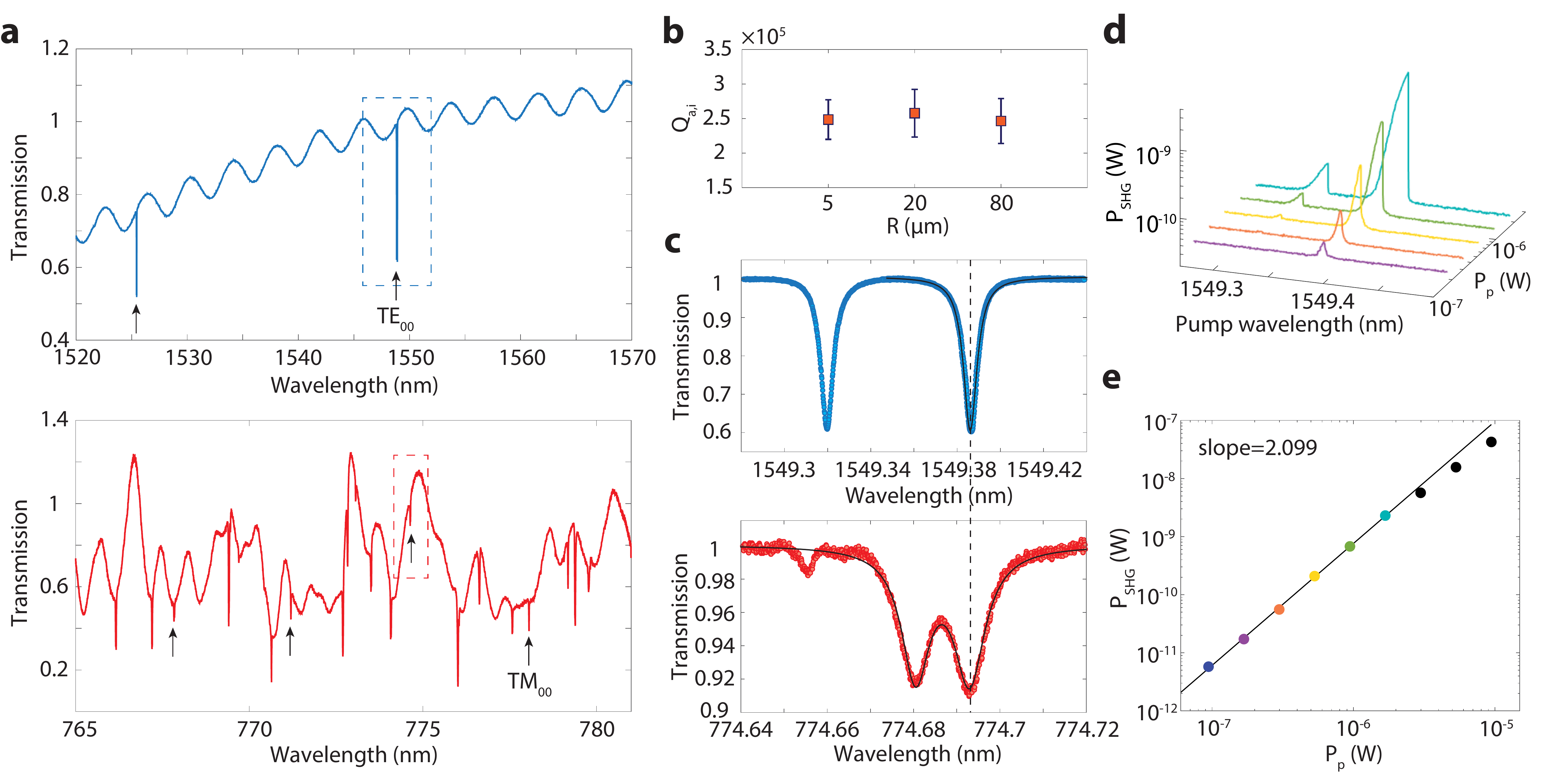}
\caption{\textbf{a}. The transmission spectrum of a 5 $\mu$m-radius microring in the 1550 nm TE and 775 nm TM band. \textbf{b}. The intrinsic quality factor of 1550 nm TE$_{00}$ resonances of microrings with different radius. The error bar represents the standard deviation of the quality factor. Eight resonances are measured for each size of the ring resonator. \textbf{c}. Normalized transmission spectrum of a pair of phase-matched fundamental and second harmonic resonances (highlighted by the dashed line), corresponding to the boxed resonances in \textbf{a}. \textbf{d}. SHG signal for various pump power (color-coded with the corresponding peak power given in \textbf{e}). \textbf{e}. Peak SHG power (subtracted with the background) versus pump power. The line fitted to the colored data points illustrates the quadratic relationship between the SHG and pump power. }
\label{fig:data}
\end{center}
\end{figure*}

\begin{figure}[!htb]
\begin{center}
\includegraphics[width=0.9\columnwidth]{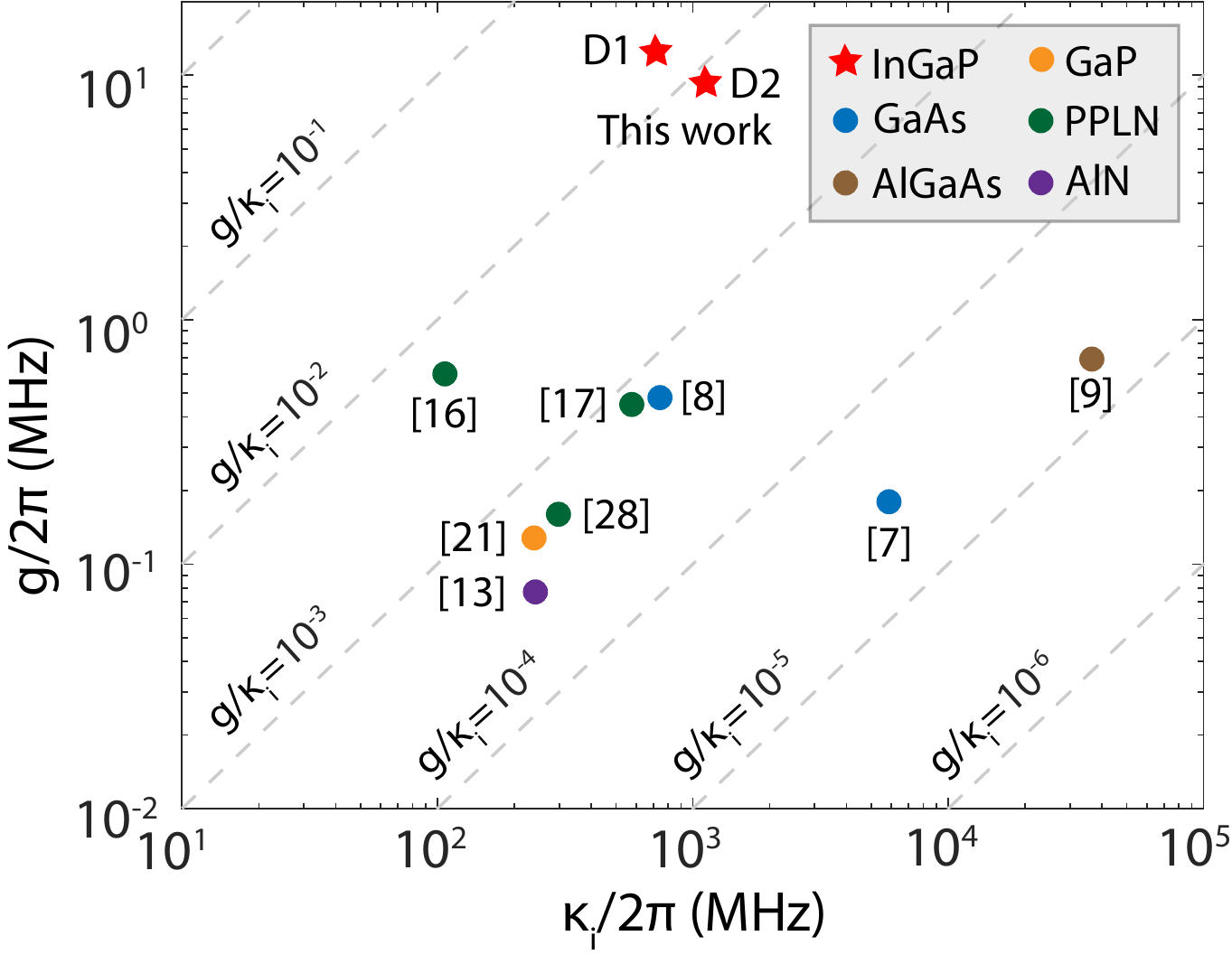}
\caption{Nonlinear mode coupling coefficient ($g$) and intrinsic photon loss rate of the fundamental mode ($\kappa_{i}$) of microring resonators made from various $\chi^{(2)}$ materials, including InGaP (this work; D1 and D2 are the devices used for SHG and SPDC experiments, respectively), GaAs (238 pm/V \cite{shoji1997absolute}) \cite{chang2019strong,kuo2014second}, Al$_x$Ga$_{1-x}$As ($<$ 238 pm/V decreasing with $x$ \cite{ohashi1993determination}) \cite{mariani2014second}, GaP (70 pm/V \cite{shoji1997absolute}) \cite{logan2018400}, periodically poled lithium niobate (PPLN) (54 pm/V) \cite{lu2020toward,ma2020ultrabright,chen2019ultra}, and AlN (1-6 pm/V) \cite{bruch201817}. The value of the dominant $\chi^{(2)}$ component of each material at 1.5 $\mu$m is indicated in parentheses. }
\label{fig:g_summary}
\end{center}
\end{figure}

\section{Device characterization}

The transmission spectrum of the microring resonator is measured using two tapered fibers evanescently coupled to the waveguide couplers. The measured coupling efficiency  is about $60\%$ and $33\%$ for the 1550 nm TE-polarized and 775 nm TM-polarized light, respectively (Appendix \ref{App:fibercoupler}). Fig. \ref{fig:data}a shows the transmission spectrum of a 5 $\mu$m-radius microring, where the 1550 nm band TE$_{00}$ and 775 nm band TM$_{00,10,20,30}$ resonances are observed and identified by their free spectral range. The quality factor of the resonance is obtained by fitting the transmission spectrum using the input-output formula. We also measured the quality factor of the 1550 nm band TE$_{00}$ resonance of 20 $\mu$m- and 80 $\mu$m-radius microrings, which is shown in Fig. \ref{fig:data}b. The similarity of the intrinsic quality factor of microrings of different sizes indicates the dominant optical loss is likely due to the surface roughness of InGaP films and/or etched sidewalls instead of waveguide bending for the ring radius down to 5 $\mu$m. The quality factor is probably not limited by material absorption because both wavelengths (1550 nm and 775 nm) are below the bandgap of InGaP and we do not observe power-dependent loss for the power the device is operated at due to nonlinear absorption, such as the two-photon absorption. However, we abstain from making smaller resonators because the gain of mode coupling will be marginal while the waveguide bending loss and the scattering loss due to sidewall roughness might become excessive.

The nonlinear mode coupling is characterized via second-harmonic generation (SHG) in quasi-phase-matched 5 $\mu$m-radius microring resonators. By sweeping the pump laser wavelength across the 1550 nm band in a group of devices with varying ring width and measuring the SHG signal power, we are able to identify quasi-phase-matched rings and resonances thereof. Finally, slight frequency mismatch between the fundamental and second-harmonic resonances could be eliminated by temperature tuning utilizing their different thermo-optic responses, which are roughly 40 pm/K and 17 pm/K for the 1550 nm TE$_{00}$ and 775 nm TM$_{00}$ resonances, respectively. The measured ring width of the quasi-phase-matched device is about 1130 nm. The discrepancy from the simulation result (945 nm) might be due to the difference of refractive index of the actual InGaP film from the value used in the simulation or the imprecision of InGaP thickness measurement (see Fig. \ref{fig:field profile&PM}c). Fig. \ref{fig:data}c shows the spectra of a pair of frequency-matched 1550 nm and 775 nm resonances satisfying the frequency- and phase-matching conditions. Due to the surface-roughness induced backscattering, these are standing-wave resonances resulted from the hybridization of the degenerate clockwise and counter-clockwise traveling-wave resonances. Using this pair of frequency-matched resonances, we measured the SHG signal with a continuous-wave pump for various powers (Fig. \ref{fig:data}d and e). The side peak in Fig. \ref{fig:data}d is from the other split fundamental resonance that is frequency-detuned from the same second-harmonic resonance. The deviation from the quadratic relationship for more intense pumps is due to thermo-optic shift of the resonances instead of pump depletion as seen from Fig. \ref{fig:data}d. However, power handling of the released device can be improved with thicker oxide cladding for better thermal conduction. The on-chip SHG efficiency is $71200\pm10300\%$/W after normalizing out the fiber-optic coupler efficiency (Appendix \ref{App:gDerivation}), where the error is the standard deviation of the measured SHG efficiency among different pump powers. We note the SHG efficiency of our device is significantly higher than a recent demonstration in thick InGaP waveguides \cite{poulvellarie2021efficient}. Together with the measured intrinsic and external quality factors of the two resonances, i.e., $Q_{a(b),i}\equiv \omega_{a(b)}/\kappa_{a(b),i}=2.63\times 10^5(9.58\times 10^4)$ and $Q_{a(b),e}\equiv \omega_{a(b)}/\kappa_{a(b),e}=9.12\times 10^5(2.12\times 10^6)$, where $\kappa_{a(b),e}$ and $\kappa_{a(b),i}$ are the external and intrinsic photon (energy) loss rate of the resonances, respectively, we extract the mode coupling coefficient $g/2\pi= 11.18\pm0.81$ MHz, which is close to the theoretical value for the device satisfying $m_b=2m_a+2$.

We use a figure-of-merit, $g/\kappa_{a,i}$, to benchmark the strength of optical nonlinearity relative to the intrinsic photon loss of the doubly-resonant $\chi^{(2)}$ cavity.  The InGaP microring resonator of this work yields $g/\kappa_{a,i}=1.52\pm0.11\%$, which is the highest among all demonstrated nonlinear photonic platforms (see Fig. \ref{fig:g_summary}). We also calculated $g^2/(\kappa_{a,i}\kappa_{b,i})$. For the InGaP SHG device it is $4.2 \times 10^{-5}$. In comparison, the LiNbO$_3$ device in \cite{lu2020toward} yields $1.9 \times 10^{-5}$. We choose $\kappa_i$ instead of the total loss rate in calculating the figure-of-merit because $\kappa_e$ might be chosen differently depending on the specific application without altering $\kappa_i$ substantially.
A majority of nonlinear optical processes, especially those relevant to quantum information applications, possess a figure-of-merit that scales with $(g/\kappa)^{2}$ (Table \ref{tab:scaling}). For example, the quantum non-demolition measurement of photons \cite{nemoto2004nearly,xia2016cavity} relies on the two-photon transport amplitude via a $\chi^{(2)}$ media which is proportional to $(g/\kappa)^{2}$ \cite{xu2015input,wang2021}. The InGaP nonlinear photonic platform is expected to significantly enhance these nonlinear optical processes for quantum information applications.

\begin{table}
\centering
 \begin{tabular}{ |c | c|}
 \hline
  $\chi^{(2)}$ nonlinear process & Scaling  \\
  \hline
  SHG efficiency & $g^2/\kappa_a^2\kappa_b$ \cite{guo2016second} \\ 
  SPDC photon-pair rate & $g^2/\kappa_a\kappa_b$ \cite{guo2017parametric} \\
  OPO pump threshold & $\kappa_a^2\kappa_b/g^2$ \cite{walls2007quantum}   \\
  Squeezed light pump & $\kappa_a^2\kappa_b/g^2$ \cite{walls2007quantum}  \\
  Two-photon transport amplitude & $g^2/\kappa_a\kappa_b$ \cite{xu2015input, wang2021} \\
 \hline
\end{tabular}
\caption{\label{tab:scaling} Scaling of the figure-of-merit of some $\chi^{(2)}$ nonlinear processes. OPO: optical parametric oscillation. Pump for squeezed light is assumed for optimal quadrature noise squeezing (i.e., $\kappa_{a,i}/\kappa_a$).  Two-photon transport is the process of two $a$-mode photons interacting via a waveguide-coupled $\chi^{(2)}$ cavity without any parametric pump, in contrast to the other four nonlinear processes. }
\end{table}

\begin{figure*}[!htb]
\begin{center}
\includegraphics[width=2\columnwidth]{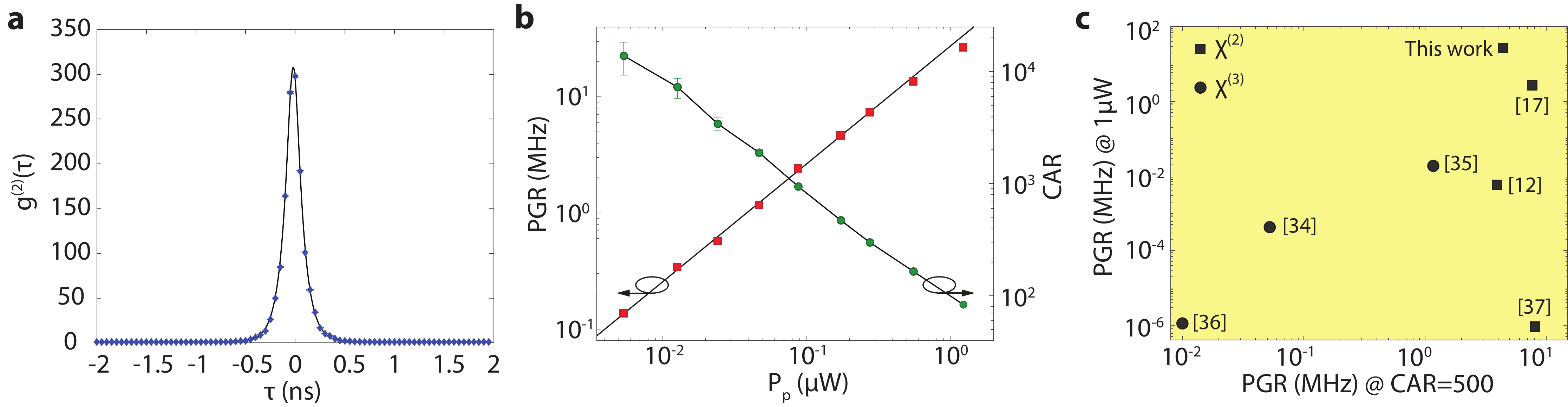}
\caption{\textbf{a}. Second-order correlation function of the SPDC photons for 0.28 $\mu$W on-chip pump power.  \textbf{b}. PGR and CAR versus pump power. The straight line is the linear fit to the PGR data with a slope of 27.5 MHz/$\mu$W. Error bar is estimated using the shot noise.  \textbf{c}. Comparison of SPDC photon-pair sources \cite{lu2019chip,ma2017silicon,wang2018photon,zhang2021,guo2017parametric,ma2020ultrabright} in terms of PGR for 1 $\mu$W pump power (i.e., efficiency) and PGR with CAR=500 (i.e., noise figure).}
\label{fig:SPDC}
\end{center}
\end{figure*}
\section{Photon-pair generation}
Here, we utilize the InGaP microring resonator with a sizable $g/\kappa$ to demonstrate highly efficient photon-pair generation via degenerate spontaneous parametric down-conversion (SPDC). For this purpose, we use a 5 $\mu$m-radius microring that is near critical coupling to both 775 nm and 1550 nm light waveguides, which is achieved by adjusting the ring-waveguide gap and waveguide wrap angle. The quality factors of a pair of phase and frequency matched resonances are $Q_{a(b),i}=1.75\times 10^5(6.18\times 10^4)$ and $Q_{a(b),e}=2.77\times 10^5(1.13\times 10^5)$. We pump the 775 nm resonance and characterize the SPDC photons from the 1550 nm resonance using two superconducting nanowire single-photon detectors (Quantum Opus, 85$\%$ efficiency, 100 Hz dark count rate) in the Hanbury Brown-Twiss setup (Fig. \ref{fig:setup}b). The SPDC signal is passed through two wavelength-division filters, each with a 40 dB extinction ratio, for filtering out the leak-through pump photons before sending to SNSPDs. The measured second-order correlation function of the SPDC photon pair shows a clear bunching effect (Fig. \ref{fig:SPDC}a). The pair-generation rate (PGR) and coincidence-to-accidental ratio (CAR) corresponding to various pump power are shown in Fig. \ref{fig:SPDC}b (see Appendix \ref{App:PhotonPair} for detailed analysis). Here, CAR is measured as $C/A-1$, where $C$ is the coincident counts at the coincidence peak and $A$ is the accidental counts estimated by averaging the background counts away from the peak. The coincidence window is chosen to be 50 ps which is close to $1/\kappa_a=87$ ps. Because the device is operated in the weak driving limit, i.e., $g\sqrt{n_p}\ll \kappa_a$ where $n_p$ is the cavity pump-photon number, the pair generation rate scales linearly with the pump power with a per-power rate of 27.5 MHz/$\mu$W which is close to the theoretical value of 41.8 MHz/$\mu$W.  The discrepancy is possibly due to slight frequency mismatch of the two resonances and pump detuning. From the photon-pair generation rate, we are able to obtain $g/\kappa_{a,i}=0.81\%$ for this device. The InGaP photon-pair source also shows an excellent CAR up to $1.4\times 10^4$. Comparing to other cavity-based SPDC photon-pair sources (Fig. \ref{fig:SPDC}c), the photon-pair generation efficiency, i.e., PGR at 1$\mu$W pump power, of our InGaP microring resonator is an order of magnitude larger than the best reported value in integrated photonic platforms \cite{ma2020ultrabright} while the noise figure is on par with typical $\chi^{(2)}$ sources. The photon-pair spectral efficiency, i.e., the per-power per-bandwidth rate, of the InGaP device is $1.5 \times 10^{4}$ W$^{-1}$ compared to $1.4 \times 10^{3}$ W$^{-1}$ of \cite{ma2020ultrabright}.  Note in principle CAR inversely scales with PGR and CAR of degenerate SPDC is half of that of non-degenerate SPDC, which explains the factor of 2 difference between the CAR of our device and that of Refs. \cite{ma2020ultrabright, zhang2021}.

\section{Discussion}
In summary, we have developed quantum nanophotonic integrated circuits in InGaP thin films with a record-breaking nonlinearity-to-loss ratio, leading to an order-of-magnitude enhancement of photon-pair generation rate compared to other nonlinear photonic platforms. The InGaP platform is expected to enhance and enable other nonlinear optical effects on chips and even observation of single-photon interactions. The oxide-on-top, suspended device architecture allows us to fabricate integrated photonic circuits in index-matched material systems without complicated wafer bonding and transfer. We point out the suspended device architecture is not fundamentally limited in terms of power handling, which can be enhanced by increasing the thickness of the oxide top-cladding.
It is promising to further improve the nonlinearity-to-loss ratio of the InGaP platform by reducing the roughness-induced photon scattering loss through, for example, tuning the III-V material growth condition \cite{xie2020ultrahigh} and optimizing the etching process.
Further, the InGaP platform is uniquely positioned, with the possibility to integrate with other III-V materials \cite{phelan2017ingap,kim1998growth}, for realizing quantum photonic microchips with integrated lasers, photodetectors, linear and nonlinear device components.

\appendix

\section{Modeling of split ring resonances}\label{App:gDerivation}
According to Ref. \cite{luo2019optical}, the $\chi^{(2)}$-mediated nonlinear mode coupling between two resonances is given by
\begin{equation}
	g  = \sqrt{\frac{\hbar\omega_a^2\omega_b}{8\epsilon_0}}\frac{\int d\mathbf{r}\chi^{(2)}_{ijk}E_{ai}^*E_{aj}^*E_{bk}}{\int d\mathbf{r} \epsilon_{\textrm{r}a}|\mathbf{E}_{a}|^2\sqrt{\int d\mathbf{r} \epsilon_{\textrm{r}b}|\mathbf{E}_b|^2}}.
\end{equation}
Applying it to the disordered InGaP with only $\chi^{(2)}_{xyz}$ component, we have
\begin{equation}\label{Eq:Appg}
	g  = \sqrt{\frac{\hbar\omega_a^2\omega_b}{8\epsilon_0}}\frac{ \int d\mathbf{r}\chi^{(2)}_{xyz}\sum\limits_{i\neq j\neq k}E_{ai}^*E_{aj}^*E_{bk}}{\int d\mathbf{r} \epsilon_{\textrm{r}a}|\mathbf{E}_{a}|^2\sqrt{\int d\mathbf{r} \epsilon_{\textrm{r}b}|\mathbf{E}_b|^2}}.
\end{equation}
To find the phase-matching condition for the traveling-wave resonances, we write the electric field in the cylindrical coordinate using the transformations
\begin{equation}
\begin{aligned}
	E_{sx} & = (E_{sr}\cos\theta-E_{s\theta}\sin\theta)e^{-im_s\theta}, \\
	E_{sy} & = (E_{sr}\sin\theta+E_{s\theta}\cos\theta)e^{-im_s\theta},
\end{aligned}
\end{equation}
where $s=a,b$. Because we choose $a$ and $b$ modes to be quasi-TE and TM polarized, we consider only the dominant terms in the numerator of Eq. \ref{Eq:Appg}, e.g.,
\begin{equation}
\begin{aligned}
	& \int d\mathbf{r}\chi^{(2)}_{xyz}E_{ax}^*E_{ay}^*E_{bz} \\
	 =& \frac{1}{4}\int d\mathbf{r} \chi^{(2)}_{xyz} \bigg[ \Big( 2E_{ar}^*E_{a\theta}^*-i(E_{ar}^{*2}-E_{a\theta}^{*2}) \Big)E_{bz} e^{i(2m_a-m_b+2)\theta} \\
	 & + \Big( 2E_{ar}^*E_{a\theta}^*+i(E_{ar}^{*2}-E_{a\theta}^{*2}) \Big)E_{bz} e^{i(2m_a-m_b-2)\theta} \bigg].
\end{aligned}
\end{equation}
This integral is nonzero only when
\begin{equation}\label{AppA5}
2m_a-m_b=\pm 2.
\end{equation}

When both modes are standing-wave resonances, i.e., even and odd mixtures of the traveling-wave resonances, their mutual coupling $g'$ can be derived as follows. Note $g$ above can be chosen to be real, thus if the second-harmonic mode is even, i.e.,
$\mathbf{E}_b^{\prime}=(\mathbf{E}_b+\mathbf{E}_b^*)/\sqrt{2}$, we have, regardless of the symmetry of the fundamental mode,
\begin{equation}
\begin{aligned}
	g^{\prime}& = \sqrt{\frac{\hbar\omega_a^2\omega_b}{8\epsilon_0}}\frac{\int d\mathbf{r}\chi^{(2)}_{ijk}E_{ai}^{\prime*}E_{aj}^{'*}E_{bk}^{\prime}}{\int d\mathbf{r} \epsilon_{\textrm{r}a}|\mathbf{E}^{\prime}_{a}|^2\sqrt{\int d\mathbf{r} \epsilon_{\textrm{r}b}|\mathbf{E}^{\prime}_b|^2}} \\
	& = \sqrt{\frac{\hbar\omega_a^2\omega_b}{64\epsilon_0}}\frac{\int d\mathbf{r}\chi^{(2)}_{ijk}(E_{ai}^{*}E_{aj}^{*}E_{bk} + E_{ai}E_{aj}E_{bk}^{*})}{\int d\mathbf{r} \epsilon_{\textrm{r}a}|\mathbf{E}_{a}|^2\sqrt{\int d\mathbf{r} \epsilon_{\textrm{r}b}|\mathbf{E}_b|^2}} \\
	& = \frac{1}{\sqrt{2}}g,
\end{aligned}
\end{equation}
where only terms satisfying Eq. \ref{AppA5} are kept. On the other hand, if the second-harmonic mode is odd, i.e., $\mathbf{E}_b^{\prime}=(\mathbf{E}_b-\mathbf{E}_b^*)/\sqrt{2}$, the coupling between the two standing-wave resonances is zero. Thus, following Ref. \cite{guo2016second}, the formula for SHG efficiency is given by Eq. \ref{eqn:SHG}, after taking into account the fact that the one-side cavity-to-waveguide loss rate is $\kappa_e/2$ for standing-wave resonances. We stress the loss rate defined in Ref. \cite{guo2016second} is half of the cavity photon (energy) loss rate used here. The normalized transmission spectrum of split resonances is fitted using the following formula to extract the loss rate:
\begin{equation}
T[\omega]=\left|1-\frac{\kappa_{e,\text{even}}/2}{\kappa_{\text{even}}/2-i(\omega-\omega_\text{even})}-\frac{\kappa_{e,\text{odd}}/2}{\kappa_{\text{odd}}/2-i(\omega-\omega_\text{odd})}\right|^2.
\label{eqn:split_trans}
\end{equation}

In the pump non-depletion region, the normalized SHG efficiency, $\eta  \equiv \frac{P_\mathrm{SHG}}{P_{p}^2}$, where $P_\mathrm{SHG}$ and $P_p$ are the on-chip SHG and pump power, respectively, for the standing-wave resonances is given by \cite{guo2016second}
\begin{equation}
	\eta=\frac{g^2}{2}\frac{\kappa_{b,e}/2}{\Delta_{b}^2+(\kappa_{b}/2)^2}\bigg(\frac{\kappa_{a,e}/2}{\Delta_{a}^2+(\kappa_{a}/2)^2}{}\bigg)^2\frac{\hbar\omega_{b}}{(\hbar\omega_{a})^2},
\label{eqn:SHG}
\end{equation}
where $\kappa_{a(b)}\equiv \kappa_{a(b),e}+\kappa_{a(b),i}$, $\kappa_{a(b),e}$  and $\kappa_{a(b),i}$ are the total, external and intrinsic photon (energy) loss rate of the resonances, respectively, $\Delta_{a}=\omega_{a}-\omega_p$, $\Delta_{b}=\omega_{b}-2\omega_p$, and $\omega_p$ is the frequency of the pump. We note the fact that the photon loss rate is roughly the same for the standing-wave and traveling-wave resonances.

\section{Device fabrication}\label{App:fabflow}
\begin{figure}[!htb]
\begin{center}
\includegraphics[width=\columnwidth]{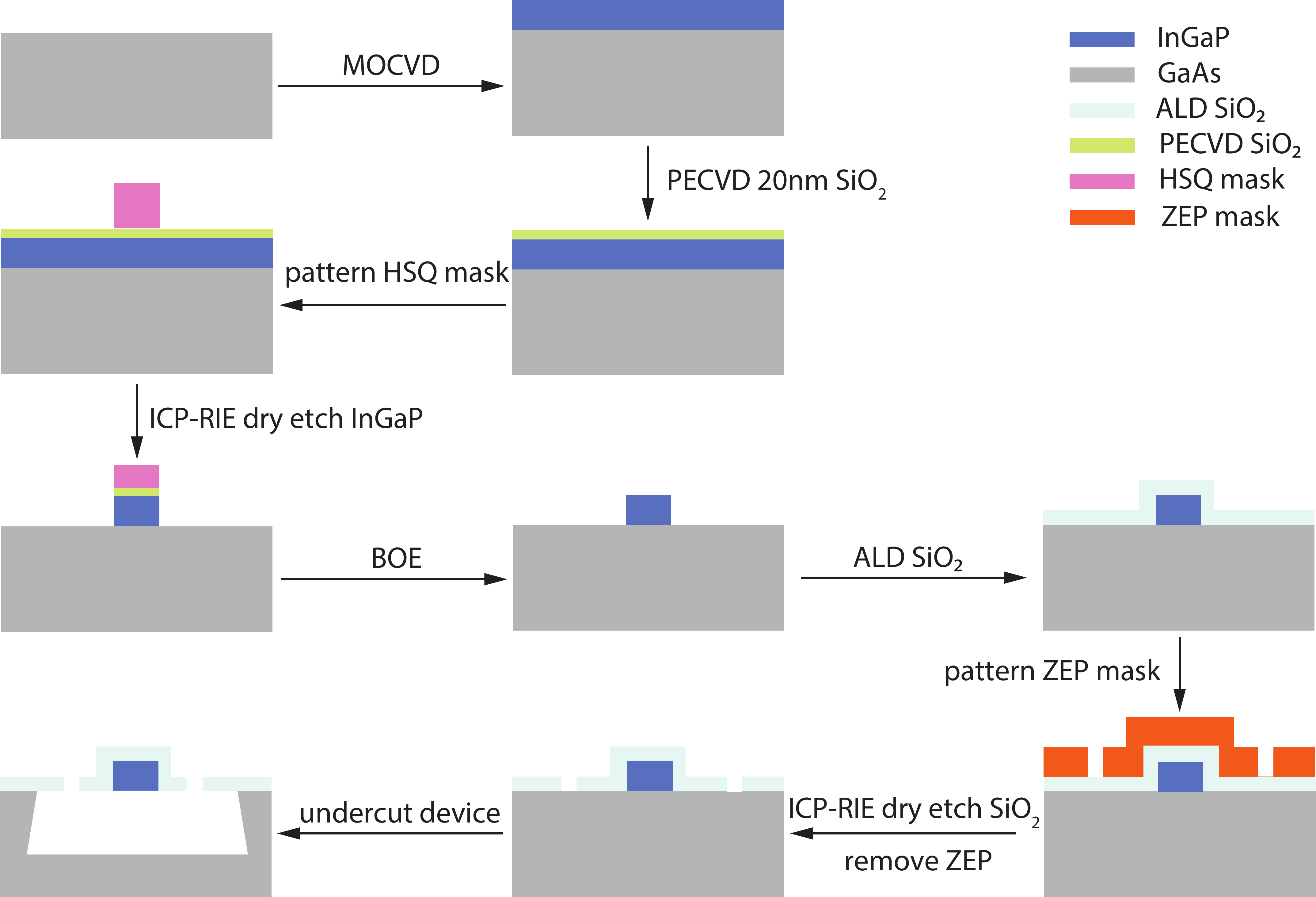}
\caption{Schematic plot of the fabrication process.}
\label{fig:flow}
\end{center}
\end{figure}

Fig. \ref{fig:flow} shows the flow chart of the fabrication process. We use citric acid-based etchant to release the InGaP device. We blend 10 g citric acid monohydrate, 20.4 mL 30\%wt hydrogen peroxide, and 47.1 mL DI water. Then we add 28 wt \% ammonia solution drops until the pH is adjusted to 8.5. The pH during the etching is monitored by a pH meter.
The released InGaP device is mechanically anchored by an oxide top cladding. In this work, we used 50 nm oxide cladding. However, thicker oxide cladding can be used for improving the power handling of the suspended devices. For example, we have tested devices with 1 $\mu$m thick oxide, which shows $\sim 20$x improvement of power handling to a level similar to silicon-on-insulator microrings, characterized by the thermo-optic shift of resonances. For the 1 $\mu$m cladding, the fabrication process needs minor adjustment by replacing the second electron beam lithography with photolithography, in order to have a sufficiently thick resist for etching through the 1 $\mu$m oxide.

\section{Fiber-optic couplers and characterization}\label{App:fibercoupler}
\begin{figure}[!htb]
\begin{center}
\includegraphics[width=0.8\columnwidth]{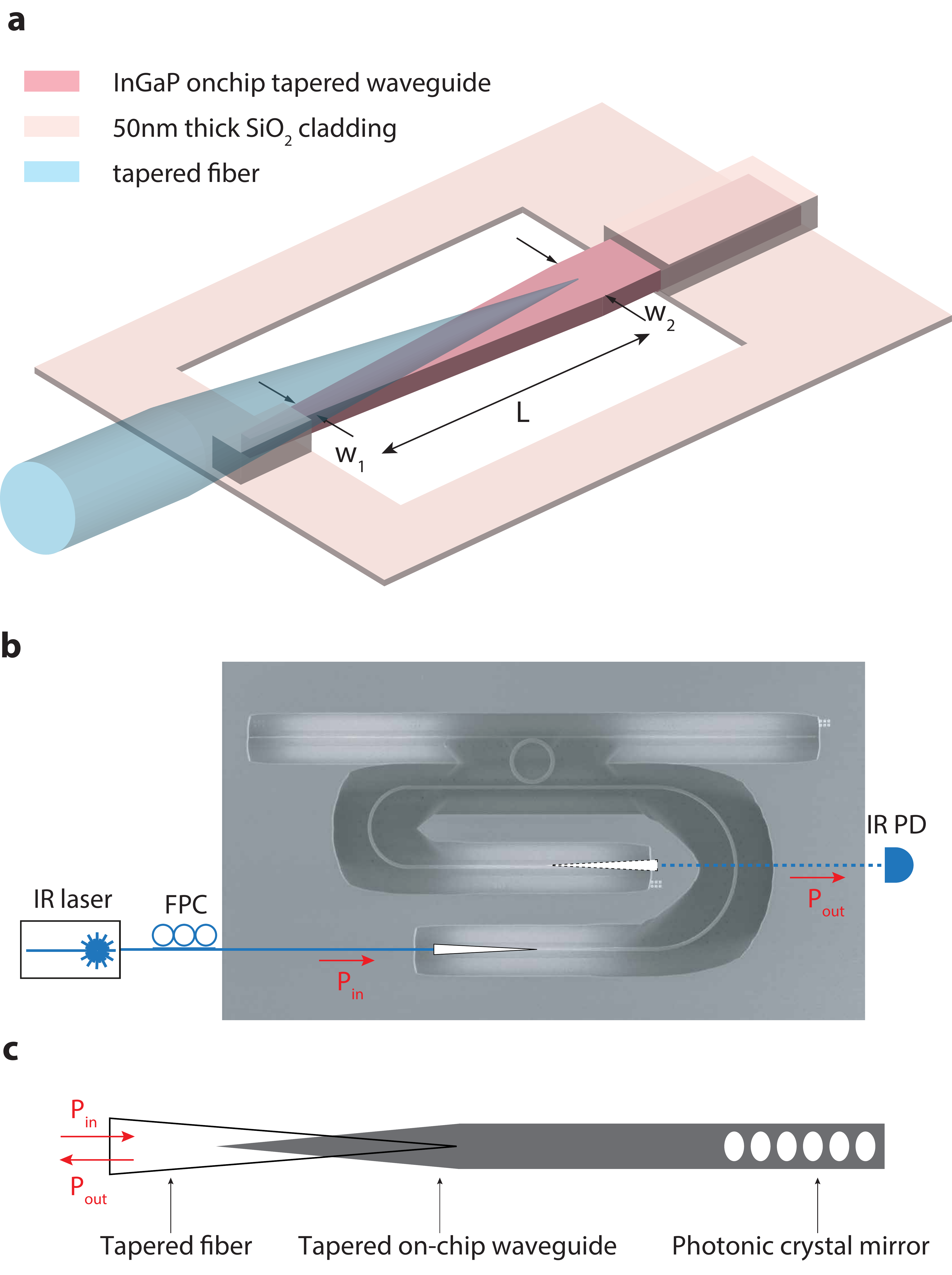}
\caption{\textbf{a}. The adiabatic fiber-optic coupler consisting of a tapered fiber  and an on-chip tapered waveguide. \textbf{b}. Transmission measurement of the device shown in the paper. \textbf{c}. Reflection measurement of a device with a photonic crystal mirror.}
\label{fig:coupler}
\end{center}
\end{figure}

As shown in Fig. \ref{fig:coupler}a, the adiabatic fiber-optic coupler consists of an on-chip tapered waveguide and a tapered optic fiber made by HF etching \cite{turner1984etch}. By controlling the fiber pulling speed during HF etching, the resulting tapered fiber has an aperture of 4\degree. The width of the tapered waveguide for coupling 1550 nm light is linearly tapered from W$_1$=50 nm to W$_2$=400 nm across a length of L=24 $\mu$m. The width of the tapered waveguide for coupling 775 nm light is linearly tapered from W$_1$=50 nm to W$_2$=300 nm across a length of L=24 $\mu$m. 

The characterization of the fiber coupling efficiency is performed as follows. There are two tapered optical fibers used in the measurement, labeled by L (left) and R (right) here. In order to characterize their individual coupling efficiency, we performed three measurements: 

(a) The transmission efficiency $\eta_1$ (Fig. \ref{fig:coupler}b; the plot shows the measurement of 1550 nm couplers). $\eta_1$ is the ratio between the output power measured right after the device and input power measured right before the device. We have $\eta_1=\eta_L\eta_R$, where $\eta_L$ and $\eta_R$ are the coupling efficiencies of the two tapered fibers. Given the short and low-loss on-chip waveguide, the propagation loss in the waveguide is ignorable.

(b,c) We also fabricated on the same chip devices consisting of the same waveguide coupler and identical photonic crystal mirrors (Fig. \ref{fig:coupler}c), and measured the reflection efficiencies $\eta_2$ and $\eta_3$ using the two fibers. We have $\eta_2=\eta_L^2 r$ and $\eta_3=\eta_R^2 r$, where $r$ is the reflection coefficient of the photonic crystal mirror. 

For all the measurements, we have adjusted the fiber position to optimize the efficiency. Then $\eta_L$ and $\eta_R$ can be calculated by
\begin{equation}
\begin{aligned}
	\eta_L &=(\eta_1^2\eta_2/\eta_3)^{1/4},
\\
	\eta_R &=(\eta_1^2\eta_3/\eta_2)^{1/4}.
\end{aligned}
\end{equation}

\section{Analysis of photon-pair generation}
\label{App:PhotonPair}
\begin{figure}[!htb]
\begin{center}
\includegraphics[width=0.9\columnwidth]{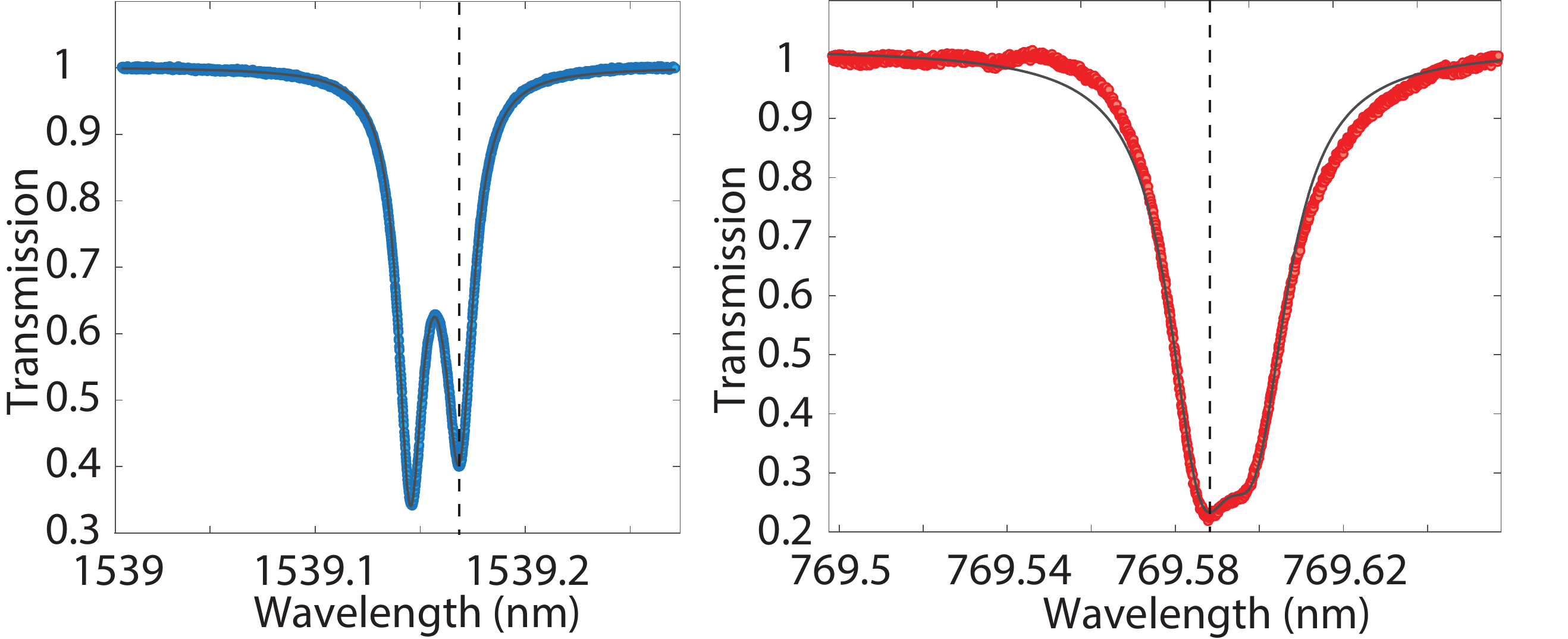}
\caption{Normalized transmission spectrum of the SPDC device. The pair of phase-matched fundamental and second-harmonic resonances are highlighted by the dashed line. Black curves are  fitting using Eq. \ref{eqn:split_trans}.}
\label{fig:resonance}
\end{center}
\end{figure}

Fig. \ref{fig:resonance} shows the transmission resonance spectra of the device used for the SPDC experiment in Section 5 of the main text.

As shown in the previous Section, only the even 775 nm resonance has a non-zero coupling with the 1550 nm resonance. Suppose the even 775 nm resonance $b_{\textrm{even}}$ couples with the even 1550 nm resonance $a_{\textrm{even}}$ because their frequency is matched. When we resonantly pump $b_{\textrm{even}}$, a pair of $a_{\textrm{even}}$ photons is generated. Because the $a_{\textrm{even}}$ mode couples bi-directionally to the waveguide, the output photon state is $\ket{1}_{R}\ket{1}_{R}+\ket{1}_{L}\ket{1}_{L}+2\ket{1}_{L}\ket{1}_{R}$, where $R$ and $L$ label the right- and left-going photons, respectively. As we detect photons from one port of the waveguide, we collect streams of both photon pairs as well as single photons.  However, only photon pairs contribute to the coincident counts, so our measurement genuinely characterized the photon-pair properties of the ring resonator, including both generation rate and coincidence-to-accidental ratio.  We stress the two-port device configuration with split resonances should not compromise the fidelity of photon pairs but only leads to a reduction of the collection efficiency by a factor of two; to restore the collection efficiency, we could simply collect photons from both ports of the waveguide or by adding a photonic crystal mirror at one end of the waveguide and thus collect all photon pairs from the other port. 

Theoretically, the total photon-pair generation rate for the microring resonator with split resonances is given by  \cite{guo2017parametric}
\begin{equation}
R=\frac{ g^{2} \kappa_{a}}{(\kappa_{a}/2)^{2}+\Delta_{a}^{2}} \frac{ \kappa_{b,e}/2}{(\kappa_{b}/2)^{2}+\Delta_{b}^{2}} \frac{P_{p}}{\hbar \omega_{b}},
\end{equation}
where $\Delta_{a}=\omega_{a}-\omega_p/2$, $\Delta_{b}=\omega_{b}-\omega_p$
The collectible, useful photon-pair rate is $\frac{\kappa_{a,e}}{\kappa_a}R$. Thus, critical coupling ($\kappa_e=\kappa_i$) for both $a$ and $b$ modes are in favor for optimizing the rate of output SPDC photon pairs.

The measured second-oder correlation function of degenerate SPDC photons is modeled by the formula \cite{guo2017parametric}
\begin{equation}\label{g2}
g^{(2)}(\tau)=1+\frac{1}{8 R \tau_{\mathrm{c}}} e^{\tau_{\mathrm{w}}^{2} / 2 \tau_{\mathrm{c}}^{2}}\left[f_{+}(\tau)+f_{-}(\tau)\right],
\end{equation}
where $f_{\pm}(\tau)=\left[1 \mp \operatorname{erf}\left(\frac{\tau \pm \tau_{\mathrm{w}}^{2} / \tau_{\mathrm{c}}}{\sqrt{2} \tau_{\mathrm{w}}}\right)\right] e^{\pm \tau / \tau_{\mathrm{c}}}$ and $\operatorname{erf}(x)$ is the error function, $R$ is the total photon-pair generation rate, $\tau_c$ is the coherence time of the SPDC photons, and $\tau_{\mathrm{w}}$ is the total timing uncertainty of the setup including detector and counter jitters and time bin width of the measurement. In our measurement, time bin width is chosen to be 50 ps. We fit the measured $g^{(2)}(\tau)$ using Eq. \ref{g2} with all parameters as fitting parameters to extract $R$. From the fitting, we also obtain $\tau_c=84$ ps, which is close to the theoretical value $1/\kappa_a=87$ ps, and $\tau_w=32$ ps.

CAR is measured as $C/A-1$, where $C$ is the coincident counts at the coincidence peak and $A$ is the accidental counts estimated by averaging the background counts away from the peak. The coincidence window is chosen to be 50 ps. The error bars are estimated using Poissonian statistics of photon counting, which means the error of photon counts is equal to the square root of the total counts, i.e., the shot noise. The error of count rate thus is given by
\begin{equation}
\label{eq:label}
\frac{\Delta R_i}{R_i}=\frac{\Delta N_i}{N_i}=1/\sqrt{R_it},
\end{equation}
where $N_i$ is the total counts, $t$ is the integration time, and $R_i=N_i/t$ is the average count rate. Error of CAR and $g^{(2)}(\tau)$ are calculated by propagating the errors of $R_1$, $R_2$ and coincidence rate $R_{12}$.


%

\vspace{2mm}
\noindent\textbf{Acknowledgements}\\ 
This work is supported by US National Science Foundation under Grant No. DMS 18-39177.

\end{document}